
\catcode`\@=11

\font\tenmsx=msxm10
\font\sevenmsx=msxm7
\font\fivemsx=msxm5
\font\tenmsy=msym10
\font\sevenmsy=msym7
\font\fivemsy=msym5
\newfam\msxfam
\newfam\msyfam
\textfont\msxfam=\tenmsx  \scriptfont\msxfam=\sevenmsx
  \scriptscriptfont\msxfam=\fivemsx
\textfont\msyfam=\tenmsy  \scriptfont\msyfam=\sevenmsy
  \scriptscriptfont\msyfam=\fivemsy

\def\hexnumber@#1{\ifnum#1<10 \number#1\else
 \ifnum#1=10 A\else\ifnum#1=11 B\else\ifnum#1=12 C\else
 \ifnum#1=13 D\else\ifnum#1=14 E\else\ifnum#1=15 F\fi\fi\fi\fi\fi\fi\fi}

\def\msx@{\hexnumber@\msxfam}
\def\msy@{\hexnumber@\msyfam}

\def\Bbb{\ifmmode\let\next\Bbb@\else
 \def\next{\errmessage{Use \string\Bbb\space only in math mode}}\fi\next}
\def\Bbb@#1{{\Bbb@@{#1}}}
\def\Bbb@@#1{\fam\msyfam#1}

\catcode`\@=\active

\font\bbigtenbf=cmbx10 scaled \magstep3
\font\twelverm=cmr10 scaled  \magstep2
\newcount\eqnno \eqnno=0
\def\numeqn{\global\advance\eqnno by 1\eqno(\the\eqnno)}
\magnification=\magstep1
\openup2\jot
\parindent 0 pt

\tolerance=1500

\def\math{\mathsurround 0pt}
\def\oversim#1#2{\lower.5pt\vbox{\baselineskip0pt \lineskip-.5pt
        \ialign{$\math#1\hfil##\hfil$\crcr#2\crcr{\scriptstyle\sim}\crcr}}}

\def\({\left(} \def\){\right)}
\def\[{\left[} \def\]{\right]}
\def\pa{\partial}
\def\frac#1#2{
  {\mathchoice{{\textstyle{#1\over #2}}}{#1\over #2}{#1\over #2}{#1 \over #2}}}

\def\unit#1{\ifinner \; 
            \else \quad \fi
            {\rm #1}}

\def\overleftrightarrow#1{\vbox{\ialign{##\crcr
    $\leftrightarrow$\crcr\noalign{\kern-1pt\nointerlineskip}
    $\hfil\displaystyle{#1}\hfil$\crcr}}}


\def\id{\Bbb 1}

\def\al{\alpha}

\def\de{\delta}
\def\ep{\epsilon}

\def\th{\theta}

\def\la{\lambda}

\def\Th{\Theta}

\def\vph{\hat\phi}
\def\F{{\cal F}}

\footline={\hfil}
\pageno=0

\hbox to \hsize{June 1993 \hfil DAMTP--93--18}
\rightline{{\tt hep-ph/9307205}}

\vskip 1 true in
\vfill
\vbox{
\centerline{\bbigtenbf The origin of the sphaleron dipole moment}
\bigskip
\bigskip
\centerline{\twelverm Mark Hindmarsh{$\,^{a)}$}}
\smallskip
\centerline{\it and}
\smallskip
\centerline{\twelverm Margaret James }
\medskip
{\it
\baselineskip=15pt
\centerline{Department of Applied Mathematics and Theoretical Physics}
\centerline{University of Cambridge}
\centerline{Silver Street}
\centerline{Cambridge CB3 9EW}
\centerline{U.K.}}}
\vskip 1 true in
\vfill
\vbox{
\baselineskip 15pt
\centerline{ABSTRACT}
By providing a suitable definition of the electromagnetic field off the
Higgs vacuum, we show that within the sphaleron there is a
monopole-antimonopole pair with quantized charges, and a loop of
electromagnetic current. On integration of the relevant charges and
currents over the interior in the limit of small $\Th_W$, we recover the
standard formula for the sphaleron dipole moment.}
{\medskip
\hrule
\smallskip\parindent 15pt\baselineskip 10pt
\item{$^{a)}$}{Present address: School of Mathematical and Physical Sciences, 
University of Sussex, Brighton, BN1 9QH, 
U.K.} }
\eject

\footline={\tenrm\hss\folio\hss}

\par {\sl 1. Introduction.} 
The sphaleron is an unstable static solution of the bosonic sector of
the standard model. Klinkhamer and Manton [1] studied the sphaleron setting the
Weinberg angle,
$\Theta_W$, to zero, and then for small $\Theta_W$ by performing a perturbation
expansion to first order in $\tan\Theta_W$. The  sphaleron of the pure SU(2)
gauge-Higgs theory is
axisymmetric and has a spherically symmetric energy density, the latter arising
from a hidden SO(4) symmetry. For non-zero Weinberg angle, 
the sphaleron has a large
magnetic moment $\mu$, which is of the order $e/\alpha_WM_{W}$,
where $\al_W = g^2/4\pi$. This is O($\al_W^{-1}$) times the dipole
moment of the $W$.  Further studies [2,3,4] have
confirmed this result and have also shown that the energy density contours
become prolate for non-zero $\Theta_W$.  

In this paper we examine the internal structure of the 
sphaleron for physical values of the Weinberg angle, in particular the way 
in which the gauge fields conspire to produce a long-range electromagnetic 
dipole field.
We build on work by Nambu [5] to show
that the sphaleron owes its dipole moment not only to a loop of electric
current  but also to a magnetic
monopole-antimonopole pair.  We are also able to use our knowledge of the inner
structure of the sphaleron to gain some qualitative understanding of its
shape at finite mixing angle.  Roughly speaking, the monopoles are joined by
a tube of Z-flux:  as $\Th_W\to \pi/2$, this tube gets thinner, resulting
in the prolate shape.  Although it is currently difficult to envisage 
experiments which probe the interior of the sphaleron, we expect  
our studies to be a useful contribution to the understanding of what we mean 
by the electromagnetic field off the vacuum.
\medskip \par
Nambu [5] developed a procedure for solving the gauge
field equations in the Higgs vacuum. In order for the
configuration to be interesting, the Higgs field must have
singularities, else it could be everywhere gauge transformed to a
constant.  The physical content of the field configuration is then
derived by examining the (possibly singular) fluxes through various
surfaces. Then one appeals to the full dynamics of the original field
equations to smear out the singularities. An essentially equivalent procedure 
was developed independently by Manton for the SO(3) Higgs theory [6]. 
\medskip \par
Nambu considered two specific cases. The first represented an isolated
magnetic monopole attached to an semi-infinite electroweak Z-string.
(The infinite electroweak Z-string was rediscovered independently and
shown to be a
solution by Vachaspati [7].) He then considered terminating the string
on an anti-monopole and conjectured that the attractive force between
the two could be balanced by spinning the two poles relative to each
other, and a quasi-stationary solution obtained.  The long range 
magnetic dipole field of this configuration resembles the sphaleron, and so  
we are
led to ask if there is a connection between this configuration and the 
sphaleron (an issue that has also been raised in references [8]). 
The answer  turns out to be a qualified yes:  in the limit 
that the length of the string tends to zero we recover the true vacuum, but   
we nevertheless find that  
part of the sphaleron  dipole moment is due to two regions of opposite magnetic 
charge.  We show that the  magnetic charge 
is partly topological in origin, for a winding number emerges when we restict 
ourselves to axisymmetric, parity
invariant configuration.  The poles do not fully explain the dipole moment, 
for we
also find a loop of electric current. We calculate the charge and current
distributions for the small $\Theta_W$ sphaleron and recover the
correct value for the dipole moment calculated in [1].
\bigskip
{\sl 2. The electroweak sphaleron.}
In the temporal
gauge the energy functional for static bosonic fields in electroweak
theory is given by
$$E=\int\left({\textstyle{1\over4}}F^a_{ij}F^a_{ij}+{\textstyle{1\over4}}f_{ij}
f_{ij}+(D_i\Phi)^{\dagger } (D_i\Phi)+\lambda
(\Phi^\dagger\Phi-{\textstyle{1\over2}}v^2)^2\right)\, d^3x,
\eqno(1a)
$$
where
$$
F^a_{ij}=\partial_iW^a_j-\partial_jW^a_i+g\epsilon^{abc}W^b_iW^c_j
\eqno(2a)
$$
$$
f_{ij}=\partial_ia_j-\partial_ja_i
\eqno(2b)
$$
$$D_i\Phi =\partial _i\Phi
-{\textstyle{1\over 2}}ig\tau^aW^a_i\Phi-{\textstyle{1\over2}}ig'a_i\Phi\
.
\eqno(2c)
$$
We have taken $T^a=-{1\over 2}i\tau^a$ as our basis for {\it L}(SU(2)) with 
$\lbrack
T^a,T^b\rbrack =\epsilon^{abc}T^c$. The Weinberg angle is given by
$\tan\Theta_W={g'/g}$.
The semiclassical masses of the W, Z, and
Higgs bosons are, respectively,
$$M_W={\textstyle1\over 2}gv,\qquad
M_Z={\textstyle1\over 2}g\sec\Th_W v,\qquad M_H=\sqrt{2\lambda}v.
\eqno(3)$$
The fields are in the Higgs vacuum if the Higgs field is on its vacuum
manifold $\Phi^{\dagger}\Phi=v^2/2$, and in addition its covariant derivative 
(2c) vanishes.
(This is distinguished from the vacuum where we would further require
$F^a_{ij}=0$ and $f_{ij}=0.$)
Following Nambu [5], we give a covariant definition of the mixing
formula, in the Higgs vacuum, which reduces to the usual formula in the unitary 
gauge (where the upper
component of the Higgs is zero and the lower component is one).  Defining
a normalised isotriplet field
$$\vph^a = {\Phi^{\dag}\tau^a\Phi/\Phi^{\dagger}\Phi} \ ,
$$
we have
$$
\eqalign{F_{ij}^{em}&=-\sin\Theta_WF^a_{ij}\vph^a+\cos\Theta_Wf_{ij}\ ,\cr
              F_{ij}^Z&=-\cos\Theta_WF^a_{ij}\vph^a-\sin\Theta_Wf_{ij}\ .\cr}
\eqno(4)$$
The field equations are
$$D_jF^a_{ij}=-{\textstyle {1\over2}}ig(\Phi^{\dagger}\tau^aD_i\Phi-(D_i
\Phi)^{\dagger}\tau^a\Phi)
\eqno(5a)$$
$$\partial_jf_{ij}=-{\textstyle{1\over2}}ig'\left(\Phi^{\dagger}D_i\Phi-(D_i
\Phi)^{\dagger}\Phi\right)
\eqno(5b)$$
$$D_iD_i\Phi=2\lambda(\Phi^{\dagger}\Phi-{\textstyle{1\over 2}}v^2)\Phi
\ .
\eqno(5c)$$
\medskip
The sphaleron solution is axially symmetric and is invariant under parity (for
details of the construction of the solution for $0\le\Theta_W\le{\pi\over 2}$
see [2]).
First we concentrate on the sphaleron at small Weinberg angle, treating it as a
perturbation of the $\Theta_W=0$ sphaleron, since the ansatz is very simple and
most of the physics is incorporated. We shall indicate the extension of our 
arguments and
results. Formally we expand in $\tan\Theta_W$ and
we give the sphaleron configuration to first order in $g'/g$ (we are taking $g$
fixed while $g'$ varies),
$$\Phi=h(\xi){v\over\sqrt2}\pmatrix{\cos\theta\cr e^{i\varphi}\sin\theta\cr},
\qquad g'a_idx^i={\textstyle 1\over 2} \left({g'\over g}\right)^2 
 p(\xi)\xi^2\sin^2\theta d\varphi$$
$$gW_idx^i=f(\xi)\pmatrix{0&e^{-i\varphi}\cr-e^{i\varphi}&0\cr}d\theta\quad+if
(\xi)
\pmatrix{\sin\theta&- \cos\theta e^{-i\varphi}\cr-
 \cos\theta e^{i\varphi}&-\sin\theta}\sin\th d\varphi\eqno(6)$$
\nobreak where $\xi=gvr.$
\par Inserting this ansatz into the field equations (5) yields the coupled
equations
$$\xi^2{d^2f\over d\xi^2}=2f(1-f)(1-2f)-{\xi^2\over 4}h^2(1-f)\eqno(7a)$$
$$\xi^2{d^2p\over d\xi^2}+4\xi{dp\over d\xi}=-h^2(1-f)\eqno(7b)$$
$${d\over d\xi}\left(\xi^2{dh\over d\xi}\right)=2h(1-f)^2+{\lambda\over
g^2}\xi^2(h^2-1)h\ ,\eqno(7c)$$ with boundary conditions $f\rightarrow
\alpha\xi^2$, $h\rightarrow \beta\xi$, and
$p\rightarrow a$ as $\xi \rightarrow 0$. As
$\xi\rightarrow \infty$,  
$p\rightarrow{b/\xi^3}$, while $f$ and $h$ tend to 1 exponentially.  The 
constants $\alpha$, $\beta$,  $a$, and $b$ are all 
O(1), and are determined on integration of the equations.
\par Asymptotically the magnetic vector potential is given, 
in the unitary gauge, by
$\vec{a}={\vec {\mu}\wedge\vec{x}/ 4\pi r^3}$,
which is a dipole field with moment $\vec{\mu}=(0,0,\mu)$ and 
strength
$$\mu={2\pi\over 3} {g'\over g^3 v}\int_0^{\infty}\xi^2h^2(\xi)(1-f(\xi))d\xi\
.\eqno(8)$$
\par\medskip
The energy density, to this order, is spherically symmetric.
Further studies [2,3,4] show that, as $\Theta_W$ increases, the energy density 
of the
sphaleron becomes prolate. Also they indicate that, for quite a large range of
$\Theta_W$ (certainly including the physical value $\sin^2\Theta_W=0.23$), the
magnetic dipole moment goes as $\mu\sim e/(\alpha_W M_W)\times constant$ where 
the
constant  depends on the Higgs mass.  For reasonable values it
is of order unity.

When $\Theta_W\ne 0$ we must relax the  symmetry restrictions and consider 
a general axisymmetric and parity invariant ansatz. 
The equatorial ($\theta=\pi/2$) slice in the gauge choice of [2] is
$$\Phi=h(\xi){v\over\sqrt2}\pmatrix{0\cr e^{i\varphi}\cr}\qquad
g'a_idx^i={\textstyle 1\over 2}\({g'\over g}\)^2 p(\xi)\xi^2d\varphi$$
$$gW_idx^i= f_1(\xi)\pmatrix{0&e^{-i\varphi}\cr
                              -e^{i\varphi} & 0 \cr}d\theta + 
if_2(\xi)\pmatrix{1&0\cr0&-1\cr}d\varphi,
\eqno(9)$$
We note that as far as the Higgs and $Z$ fields are concerned 
this is also the ansatz for the electroweak Z-string [6,7].
There is also a superficial resemblance to the monopole-antimonopole
configurations studied by Nambu. \bigskip \par

{\sl 3. Monopole-antimonopole configurations.}
In [5] Nambu developed a procedure for  solving the
gauge field equations in regions where the Higgs field is in the vacuum. With
his procedure, the configuration is only interesting if $\Phi$ is singular
somewhere, else it is gauge equivalent to the vacuum. Correspondingly, the
gauge fields will also have singularities where the Higgs field vanishes. The
true solution of the constrained field equations (a constraint must be imposed
to ensure that $\Phi$ continues to vanish on the singular points), smooths 
out the singularities. However, as Nambu asserted, much of the interesting
physics, including what the configuration actually represents e.g. monopole,
string can be deduced from the singular case by the computation of fluxes
through suitably chosen surfaces.  Nambu's method is actually a way of 
choosing the gauge in the electroweak Higgs vacuum,
similar to one developed by Manton for SO(3) [6]. 

\medskip \par The procedure is to solve
$$D_i\Phi=0\eqno(10)$$ 
to obtain expressions for the gauge potentials in terms 
of the Higgs field.    The result is
$$gW_i^a=-\epsilon^{abc}\vph^b\partial_i\vph^c
-i\chi\vph^a(\Phi^{\dagger}\partial_i\Phi-
\partial_i\Phi^{\dagger}\Phi)-g\sin\Theta_W \vph^a\alpha_i\ ,$$
$$g'a_i=-i\eta(\Phi^{\dagger}\partial_i\Phi-
\partial_i\Phi^{\dagger}\Phi) + g'\cos\Theta_W\al_i\eqno(11)$$ where 
$\chi+\eta=1$. The vector function $\al_i$ is  undetermined, for 
the abelian part of the gauge field $F_{ij}^{em}$ is not 
fixed by $D_i\Phi=0$ alone.  Following 
Manton [6] it is possible to show that setting $\al_i =0$ is a gauge choice 
in which the isovector Higgs field $\vph^a$ is constant along the lines of 
electromagnetic flux computed from the potentials (11): that is, 
$B_i^{em}\pa_i\vph^a =0$.  
In this gauge we cannot write down an arbitrary form 
for the Higgs field, for the resulting electromagnetic field
must satisfy the vacuum  Maxwell's equations.
\medskip
Nambu considered first the configuration
$$\Phi=\pmatrix{\cos{\theta\over
2}\cr\sin{\theta\over 2}e^{i\varphi}\cr},
\eqno(12)$$
which is singular on the line $\th=\pi$.  The resulting gauge potentials are
$$
\eqalign{
gW_i^a &= \ep^{aij}x_j/r^2 + \chi \ep^{ij3}x_ax_j/r^2(r+z) \cr
g'a_i &= -\eta\ep^{ij3}x_j/r(r+z)\cr}\eqno(13)
$$
If one computes the SU(2) flux $F_{ij}^a\vph^a$ and the U(1) flux $f_{ij}$, 
both out of the origin and along the singularity, one finds that
the SU(2) gauge field is that of a  magnetic monopole of charge
$-{4\pi/ g}$. A portion, $-{4\pi\eta/ g}$, of the flux spreads out and
the rest, $-{4\pi\chi/ g}$, is confined to the tube.
The U(1) field, while sourceless, gives rise to a spreading
monopole field, the flux ${4\pi\eta/ g'}$ being precisely compensated by
the returned flux ${4\pi\eta/ g'}$ in the string. The net effect is a
spreading field due to a magnetic charge $Q={4\pi\eta/ e}$, and a flux tube
on the negative $z$-axis containing a mixture of an electromagnetic flux and a 
quantised Z-flux, given respectively by
$${4\pi\over
e}(-\chi\sin^2\Theta_W+\eta\cos^2\Theta_W),\qquad\-{4\pi\over
e}\sin\Theta_W\cos\Theta_W.\eqno(14)$$
 Note that in order to compute the flux along the singularity,
Nambu uses Stokes' theorem and hence implicitly assumes the abelianization of
the field tensors within the singularity, i.e., where the
fields are off the Higgs vacuum. This assumption is commonly made in the
literature for strings, and we shall see its importance later.
 The Z-flux is quantised as a result of the $2\pi$ phase change of the
Higgs field around the string, while the electromagnetic flux in the tube
is taken to be zero  (thus the 
singular line is essentially an
 electroweak string [7]).  Therefore the condition that there is no 
electromagnetic flux in line singularities fixes 
$\eta=\sin^2\Theta_W$.  Hence we have a monopole with magnetic charge
$$Q={4\pi\over e}\sin^2\Theta_W={e\over \alpha_W} \eqno(15)$$ and a Z-string on
the negative z-axis. This configuration has infinite energy, but Nambu
envisaged constructing finite energy configurations by taking a
monopole-antimonopole pair of this charge separated by a finite 
length of Z-string. He
then conjectured that a quasi-stationary solution to the bosonic sector of the
Standard Model could be constructed by spinning this `dumb-bell' so that the
string tension would be counteracted by the centrifugal force on the poles. We
have in mind a different possibility, for if the poles
were separated by a distance $M^{-1}_W$, the natural length scale of the
problem, it would then have a dipole moment of the same order and with the same
$\Theta_W$ dependence as the sphaleron.  This leads us to investigate whether 
the sphaleron can be viewed as a monopole-antimonopole pair.

\medskip \par In Nambu's gauge the 
singular Higgs field representing a monopole-antimonopole pair situated on the
$z$ axis at $z=\pm d $, and connected by a string, is
$$\Phi=\pmatrix{\cos{\Theta}\cr\sin{\Theta}e^{i\varphi}\cr}\eqno(16)$$ where
$\cos 2\Theta=\cos\theta_+-\cos\theta_- +1$, and $\theta_{\pm}$ 
is the polar angle seen from
the poles [8]. The field is singular on the line joining the  two points. 
 The fluxes,
both SU(2) and U(1), out of the northern and southern hemispheres at infinity
are zero. The total U(1) flux through the equatorial plane is zero. The SU(2)
flux downward through the equatorial plane is $-{4\pi/ g}$, and so the 
electromagnetic flux is $4\pi\sin\Theta_W/g$ as expected.   However, in 
the limit $d\to 0$ where the singular line reduces to a point we recover the 
vacuum ($\Phi = (1,0)^T$, $W_i^a = 0 = a_i$), 
rather than the sphaleron, which has a long range dipole field.   
Nevertheless, we are still prompted to ask if there are any 
monopoles lurking in the sphaleron supplying its dipole moment. 
To answer the question we must equip ourselves with definitions of the
electromagnetic and Z field tensors in the interior of the sphaleron, where the
fields are not in the Higgs vacuum. 

\bigskip {\sl 4. The origin of the
sphaleron dipole moment.} 
Coleman [9] makes it clear that there
is no unambiguous definition of the electromagnetic field off the Higgs vacuum.
Different choices correspond to different idealised magnetometers. Hence we are
free to make our choice subject to the constraint that on the Higgs vacuum it
reduces to (4) and that the field tensor is gauge invariant. The choice we make 
is that (4) applies everywhere except where the Higgs field vanishes. 
This 
appears to us to be natural; it gives rise to simple formulae and the physics
involves monopoles of quantised charge. 't Hooft [10] gives a different
definition with a non-physical
singularity at the origin, which we shall discuss in the last section. 
\par The usual mixing formula arises from
diagonalizing the mass matrix for the gauge fields in the background of the
Higgs field, on the vacuum manifold, in the unitary gauge. The electromagnetic
field is then associated with the massless field. However we may equally well
perform such a diagonalization in the background of any non-vanishing $\Phi$.
The relevant terms, arising from the $(D_{\mu}\Phi)^{\dagger}(D^{\mu}\Phi)$
term in the lagrangian, are 
$$\Phi^{\dagger}\Phi\left(g^2W_i^a W_i^a+2gg'W_i^a
a_i\vph^a+g'^2a_i\right). \eqno(17) $$
Equivalently we can note that for any non-zero $\Phi$,
$({\id}+\vph^a\tau^a)$  
generates a U(1) symmetry which leaves $\Phi$ invariant. If we define 
the massive Z-field
and massless electromagnetic field by 
$$
\eqalign{
Z_i & = -\cos\Theta_W W_i^a\vph^a - \sin\Theta_W a_i \cr
A_i &= -\sin\Theta_W W_i^a\vph^a + \cos\Theta_W a_i\cr}
\eqno(18)
$$
then the gauge invariant field strength tensors are given for all $\Phi 
\ne 0$ by
$$F_{ij}^Z=\partial_iZ_j-\partial_jZ_i - {1\over g}\cos\Theta_W(
\vph^a\partial_i\vph^b\partial_j\vph^c - \vph^a D_i\vph^bD_j\vph^c)
\epsilon^{abc},\eqno(19a)$$
$$F_{ij}^{em}=\partial_iA_j-\partial_jA_i- {1\over g}\sin\Theta_W(
\vph^a\partial_i\vph^b\partial_j\vph^c - \vph^a D_i\vph^bD_j\vph^c)
\epsilon^{abc}.\eqno(19b)$$
We see explicitly that away from the Higgs vacuum 
the fields are
non-abelian due to the SU(2) contribution. There, the usual Maxwell 
equations are no
longer satisfied. Defining the SU(2) magnetic field
$$B_i^{SU(2)}={\textstyle{1\over 2}}\epsilon_{ijk}(F^a_{jk}\vph^a)\,\eqno(20)$$
then $$\nabla \cdot B^{SU(2)}\ = {\textstyle{1\over
2}}\epsilon_{ijk}F^a_{jk}D_i\vph^a\ =\ \rho_m$$ is an SU(2) magnetic
charge density which will in general be non-vanishing. Via the mixing formula an
{\it electromagnetic} magnetic charge density is obtained,
$$\rho_m^{em}= -\sin\Theta_W\rho_m.\eqno(21)$$ Using the field equations (5) we
obtain the electromagnetic current
$$J_i^{em}=\partial_jF_{ij}^{em}= -\sin\Theta_WF_{ij}^a
D_j\vph^a.\eqno(22)$$   
The volume integral over $z\geq0$ of the charge (21) may be written via Gauss's
theorem as the integral of the related magnetic field (20) over the northern
hemisphere at infinity plus the integral over the equatorial plane.
Asymptotically the the gauge field is a dipole and hence the integral
over the hemisphere vanishes. For the integral over the plane,  we find that 
the field abelianises there, and so
$$B_i^{SU(2)}= -(\nabla\wedge W^3)_i\ .\eqno(23)$$ 
Hence using Stokes' theorem the integral is
written as the integral of $W_i^3$ round the circle at infinity, which is just
${4\pi/ g}$ times the winding number of the lower component of the Higgs, 
which is 1.  Repeating the procedure in the lower hemisphere results 
in the integration 
being taken the other way around the circle.  
The fact that we obtain the `correct'
quantized value $-{4\pi/ g}$ for the total SU(2) charge in $z\geq0$ (and
$+{4\pi/ g}$ for $z\leq0$) indicates underlying topological considerations.
In deriving this result we have assumed nothing except that the field 
configuration is axisymmetric and parity 
invariant, so that the equatorial slice has the form (9) [2]. This 
class of configurations encompasses the sphaleron at all values of
$\Th_W$.  We have found that for all such configurations the magnetic charge is 
quantized:  it is in this restricted sense that the charge is topological. 
The integral of $\rho_m^{em}$ in the upper (lower) hemisphere of the 
sphaleron is $-(+)4\pi\sin\Theta_W/g$. 
Recalling Nambu's model of the isolated monopole and string we note that the
implicit assumption of the abelianization of the fields off the Higgs vacuum 
is essential in using Stokes' theorem to compute the fluxes. 

\medskip \par We know the total charge in the lower and upper hemispheres for
the sphaleron but we need to find the charge distribution and the current (22)
in order to calculate the magnetic dipole moment. Again, we shall work with the
small $\Theta_W$ configuration (6). We obtain for the magnetic charge density,
$\rho_{em}$, and the electromagnetic current (to first order in $g'/ g$)
$$\rho_{em}=8g'\cos\theta {df\over dr}{(1-f)\over g^2r^2},\eqno(24a)$$
$$\vec{J}^{em}=8g'{f(f-1)^2\over g^2r^4}(-y,x,0).\eqno(24b)$$ As a check,
integrating the magnetic charge density over the northern (southern) hemisphere
gives $\pm {4\pi g'/ g^2}$ as expected. 
(In Figure 1 we exhibit these charge and current densities, using a
numerical computation of the function $f$ [4].)
The definition of the dipole moment is $$\vec {\mu}=\int\left(\vec{x}\rho^{em}\
+\ {\textstyle 1\over 2}\vec{x}\wedge\vec{J}^{em} \right)d^3x\ . \eqno(25)$$ On
substituting (24) we find $\vec{\mu}=(0,0,\mu)$ where the strength $\mu$ is
given by $$\mu={32\pi g'\over 3g^2}\int_0^{\infty} \left[r{df\over dr}(1-f)+
f(1-f)^2 \right]dr\ .\eqno(26)$$ 
The charge and the current contribute 70\% and 30\% respectively.  
Furthermore, on integration by parts and using the field equations, we
recover, after a little  algebra, expression (8) for $\mu$.

This result lends weight to our definitions of $\rho_m^{em}$
and $\vec{J}^{em}$, and with them we are able to build a physical picture
of the source of the sphaleron dipole moment.  The sphaleron contains
a monopole-antimonopole pair  encircled
by a ring of current.  The monopoles are joined by a short tube of 
Z flux, for in the equatorial plane 
$$
B_i^Z = -{2\over g r}{df\over dr} \de_i^3,
\eqno(27)
$$
which is in the $-z$ direction, peaked at the origin, and decays 
exponentially away from it.

{\sl 5. Discussion. }
We can use our new picture of the sphaleron to gain a qualitative
understanding of the sphaleron in the $\Th_W\to\pi/2$ limit, which is
problematic for  numerical approaches [2].  There are various ways of taking
this limit: if we fix $g$ then we may keep fixed either $g'/\sqrt{\la}$
or $g/\sqrt{\la}$.

Let us consider first taking the Higgs mass to infinity
with the Z mass.  We have already noted that the sphaleron ansatz
resembles the electroweak string in the equatorial plane, and that
the gauge field abelianizes there.  Thus we expect that the width of
the tube of Z flux and of the region where the Higgs leaves its vacuum
manifold should behave in much the same way as for the string.  As we take
$M_H$ and $M_Z$ to infinity the width of the flux tube should shrink
to zero (relative to $M_W$).  We conjecture that $M_W$ sets the scale
of the size of the monopoles, and thus for the size of the sphaleron
in the $z$ direction.  Thus in the $\Th_W\to\pi/2$ limit with $M_Z/M_H$
fixed we expect the sphaleron to become extremely prolate.  Indeed, this limit
is equivalent to taking $g\to 0$, with $g'$ and $\la$ fixed, in which we
obtain a theory with global symmetries, which has vortex solutions known as
semilocal strings [11].  Thus we can say that in this limit the sphaleron turns
into a segment of semilocal string. If instead we take $g'$ to infinity while
keeping the Higgs mass fixed, then the width of the flux tube should again
decrease, albeit much more slowly [12].

If our picture of the magnetic dipole moment is correct, we should expect
that it remains proportional to $4\pi\sin\Th_W/gM_W$,
although the shape distortion
will change the overall constant.  Indeed, to order $(g'/g)^3$ it is found [4] 
that
$$
\mu \simeq {4\pi\over gM_W} 7.0\left({g'\over g} - 0.42\left({g'\over 
g}\right)^3\right).
\eqno(28)
$$
The value for the coefficient of the cubic term compares reasonably well
with the expected value 0.5. Numerically [2], it seems that  the overall
constant increases somewhat as $g'$ is taken to infinity with the Higgs mass
fixed.  A possible explanation is that the stronger electromagnetic
interactions between the poles and the loop are forcing them further apart.
\medskip
Our definition of the electromagnetic field off the vacuum is not the only one: 
't Hooft proposed that we take
$$
\eqalign{
\F^{em}_{ij} &= F^{em}_{ij} - {1\over g} \sin \Theta_W\vph^aD_i\vph^bD_j\vph^c
\epsilon^{abc},\cr
&= \partial_iA_j - \partial_jA_i - 
{1\over g}\sin\Theta_W\vph^a\partial_i\vph^b\partial_j\vph^c
\epsilon^{abc}.\cr}
\eqno(29)
$$
In the unitary gauge, where $\vph^a$ is constant, this reduces to an appealing 
form, that of ordinary electromagnetism, which of course has identically zero   
magnetic charge.  However, for the sphaleron we find that we have to face an 
$O(r^{-2})$ singularity in the magnetic field at the origin, which is not 
so attractive.  It seems better to us to keep physical fields bounded 
in smooth configurations like the sphaleron: hence our preference for (4).

\bigskip
We are indebted to Nick Manton for many helpful discussions, and in 
particular for clarifying the issue of the gauge 
choice for the monopole-antimonopole system and its distinction from 
the sphaleron.  
We thank also Stuart Rankin, Ana Ach\'ucarro and Anne-Christine Davis. 
This work is funded by SERC.

\vfill\break
{\sl References} \medskip
{\parindent 25pt{\frenchspacing
\item{[1]} F. Klinkhamer and N. Manton, Phys. Rev. D30, 2212 (1984).
\item{[2]} B. Kleihaus, J. Kunz and Y. Brihaye, Phys. Lett. B273, 100 (1991);
B. Kleihaus, J. Kunz and Y. Brihaye, Phys. Rev. D46, 3587 (1992).
\item{[3]} F. Klinkhamer, R. Laterveer, Z. Phys. C 53, 247 (1992).
\item{[4]} M. James, Z. Phys. C 55, 515 (1992).
\item{[5]} Y. Nambu, Nucl. Phys. B130, 505 (1977).
\item{[6]} N. Manton, Nucl. Phys. B126, 525 (1977).
\item{[7]} T. Vachaspati, Phys. Rev. Lett. 68, 1977 (1992).
\item{[8]} T. Vachaspati, Tufts University preprint TUTP-92-12 (1992); 
T. Vachaspati and M. Barriola, Tufts University preprint TUTP-93-?? (1993).\par}
\item{[9]} S. Coleman, `Aspects of Symmetry', Cambridge University Press,
Cambridge (1985).
{\frenchspacing
\item{[10]} G. 't Hooft, Nucl. Phys. B79, 276 (1974).
\item{[11]} T. Vachaspati and A. Ach\'ucarro, Phys. Rev. D44, 3067 (1991);
M.Hindmarsh, Phys. Rev. Lett. 68, 1263 (1992).
\item{[12]} E. B. Bogomol'nyi and A. Vainshtein, Sov. J. Nucl. Phys. 23, 588
 (1976); C. T. Hill, H. M. Hodges and M. Turner, Phys. Rev. Lett. 59, 2493
 (1987).\par}
}

\vfill\eject
{\bf FIGURE CAPTIONS}
\medskip
{\sl Figure 1a.} 
The surface enclosing 75\% of the magnetic charge of the sphaleron, 
in the limit 
of small Weinberg angle.  The roughness of the surface is a numerical 
artefact.  Distances are measured in units of $\xi = 2M_Wr$, where $M_W$ is 
the mass of the $W$.
\smallskip
{\sl Figure 1b.}
The surface enclosing 75\% of the electric current in the sphaleron, again 
in the small $\Theta_W$ limit.
\vfill\eject

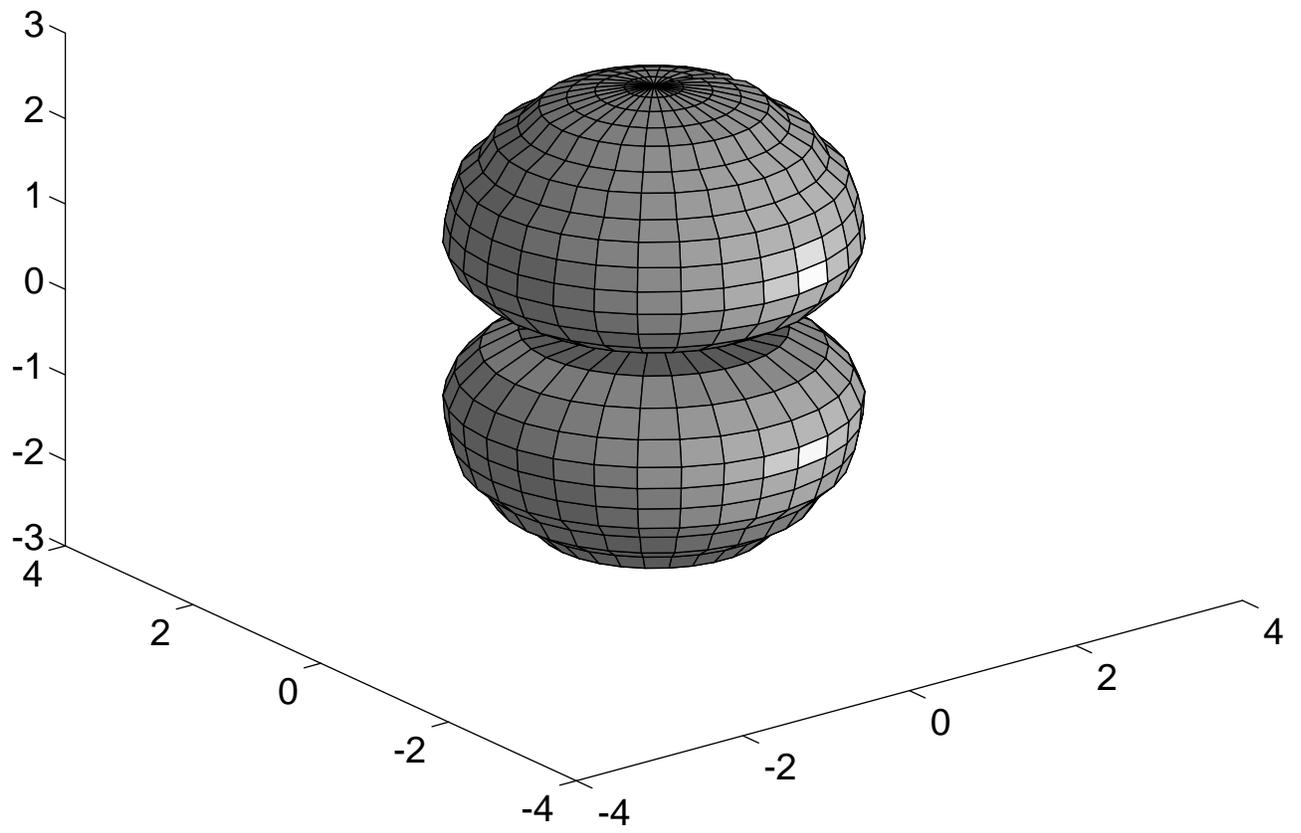
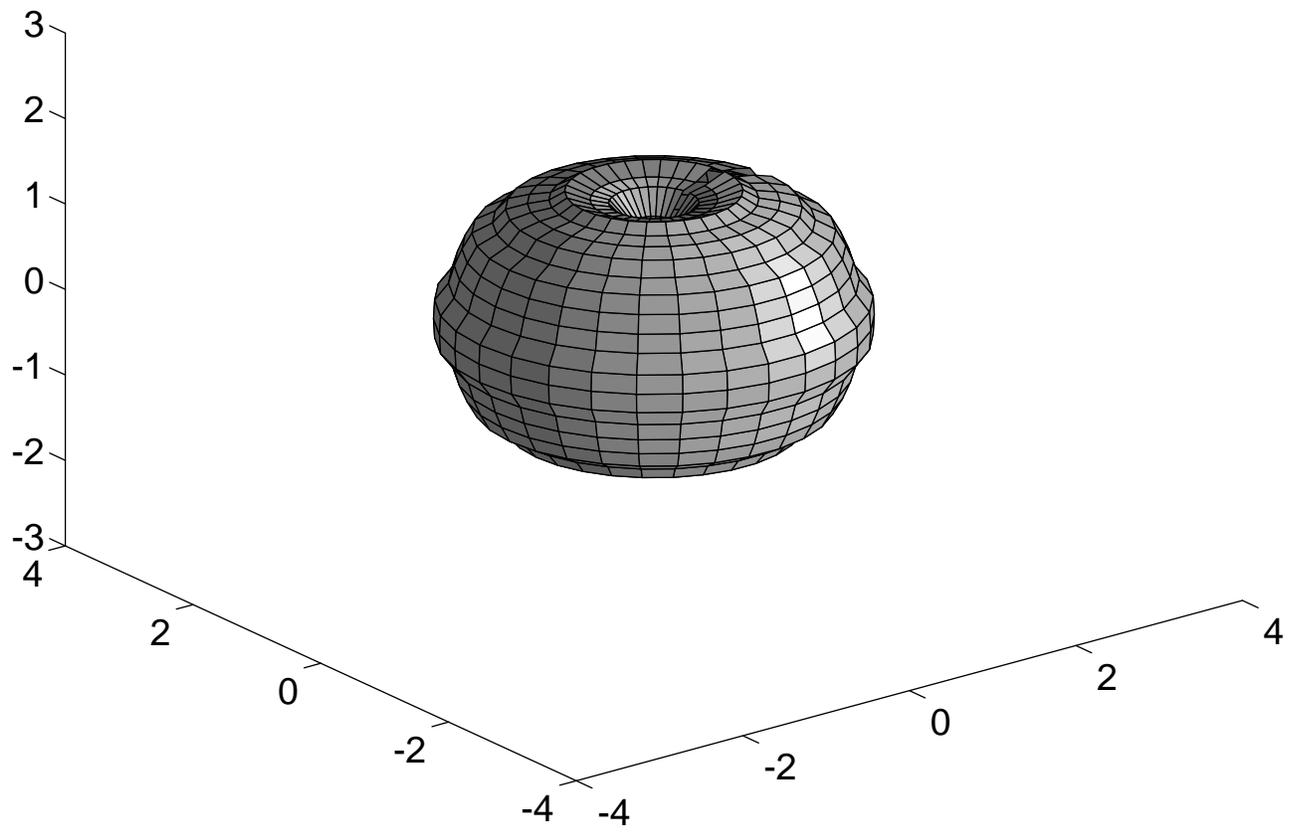
\end